\documentclass[twocolumn,preprintnumbers,amsmath,amssymb]{revtex4-1}
\usepackage{graphicx}
\usepackage{dcolumn}
\usepackage{amsmath,amsbsy,amssymb,scalefnt}
\usepackage{bm}
\usepackage{epstopdf}
\usepackage{amsfonts}
\usepackage{textcomp}
\usepackage{mathtools}
\usepackage{hyperref}
\usepackage{color,soul}
\usepackage[normalem]{ulem}
\newcommand{\stkout}[1]{\ifmmode\text{\sout{\ensuremath{#1}}}\else\sout{#1}\fi}
\newcommand\rst{\bgroup\markoverwith
{\textcolor{red}{\rule[.5ex]{2pt}{1pt}}}\ULon}
\setstcolor{red}
\hypersetup{backref=true,
 pdfnewwindow=true, colorlinks=true,
 linkcolor=blue, anchorcolor=blue,
 citecolor=blue, filecolor=blue,
 menucolor=blue, urlcolor=blue}

\begin{document}


\title{Enhanced thermoelectric properties in phosphorene nanorings}

\author{Fatemeh Moghadasi Borojeni$^1$}
\author{Esmaeil Taghizadeh Sisakht$^{1}$}
\author{Farhad Fazileh$^{1,}$}
\email{fazileh@iut.ac.ir}
\author{F. M. Peeters$^{2,3}$}
\affiliation{$^1$Department of Physics, Isfahan University of Technology, Isfahan 84156-83111, Iran}
\affiliation{$^2$HSE University, Moscow, 10100 Russia}
\affiliation{$^3$Departamento de Fisica, Universidade Federal do Ceara, 60455-760 Fortaleza, Ceara, Brazil}

\date{\today}

\begin{abstract}
Using the tight-binding approach, we investigate  the thermoelectric properties of rectangular phosphorene nanorings for both symmetrically and asymmetrically attaching to phosphorene nanoribbon leads. We design our phosphorene-based nanostructures to enhance the thermoelectric performance in the absence and the presence of perpendicular magnetic fields. Our results show that when zigzag phosphorene nanoribbons (ZPNRs) 
are coupled symmetrically to rectangular rings, a comparatively large band gap is induced in the electronic conductance due to the suppression of the contribution of edge states. This gives rise to a remarkable increase in the thermopower response compared to the case of pristine ZPNRs. More intriguingly, we found that though the maximum power factor in this system is about the same as the one for its ZPNR counterpart, the much smaller electronic thermal conductance of this phosphorene-based nanostructure can remarkably contribute to the improvement of the figure of merit. Also, we found that the symmetry/asymmetry of our designed nanostructures, the geometrical characteristics of the ring, and the magnetic flux are three important factors that control the thermoelectric properties of phosphorene quantum rings.
Our numerical calculations show that by changing the magnetic flux through the nanoring, a drastic increase in the thermopower is observed near an antiresonance point. We demonstrate the tunability of the thermopower and the possibility to switch on and off the thermoelectric response of phosphorene nanorings with the magnetic flux. Moreover, for asymmetric connection configurations with armchair-edged leads, we found that though the thermopower is almost intact, a remarkable reduction of the electronic thermal conductance can lead to a notable improvement in the figure of merit. Our results suggest phosphorene nanorings as promising candidate nanostructures for thermoelectric applications.
\end{abstract}

\maketitle


\section{\label{introduction}Introduction}
Designing devices with optimal thermoelectric (TE) properties is highly desirable for the future of energy harvesting and environmental issues~\cite{n47}.
TE materials occupy a special place in clean energy research field, since they 
directly convert heat into electrical energy and vice versa. They are promising candidate materials for the future of clean and efficient electrical power generators and cooling (heating) devices. 
The TE efficiency of a material is often assessed by dimensionless figure of merit $ZT=\frac{S^2 \sigma T}{\kappa}$; where $S$ is the Seebeck coefficient (SC) or thermopower, $\sigma$ and $\kappa$ are the electrical and thermal conductivities, respectively and $T$ is the temperature. 
This relation suggests that a reduction in thermal conductivity (while the electrical conductivity is kept almost the same) leads to a strong improvement of the TE efficiency. 
One way to accomplish this task is to make nanostructures out of the bulk TE materials~\cite{n48}.
Here, the enhanced TE factor of merit $ZT$ is due to phonon scattering with boundaries and quantum confinement effects. The TE efficiency can be further enhanced by making nanorings out of the materials~\cite{n28,n29}. 
Using nanorings made of nanoribbons of the pristine materials we can take advantage of the Aharonov-Bohm (AB) effect when we apply an external magnetic field to tune the magnetoresistance of the system~\cite{n44} and further enhance the TE efficiency~\cite{n30,n31}.

Among nanostructured semiconducting materials, two dimensional (2D) phosphorene has recently received remarkable attention
due to its great transport properties and potential applications~\cite{n1,n2,n3,n4,n5,n6,n7,n8}. 
This structure consists of a single- or few-layer of black phosphorus and has been successfully fabricated by researchers~\cite{n3,n4,n7}.  From these studies, it is suggested that because of its remarkable electronic, mechanical, and optical  properties, it 
offers great promise for applications in electronic and optoelectronic devices~\cite{n9,n10,n11,n12,n13,n14,n15,n16,n17,n18}. 
In addition, due to superior thermoelectric characteristics,
it is also proposed that phosphorene nanostructures are highly promised for thermoelectric devices and it could be their key application in the future~\cite{n19}. 
Various factors may significantly affect the three parameters that are used to quantify the figure of merit in this structure.  For single-crystals of bulk black phosphorus,  the experimental measurements revealed that the value of SC is about 340~$\mu$V/K at room temperature~\cite{n20}.   It was shown that gate-tuning is a successful way to control the TE power coefficient 
 in a thin flake of black phosphorus~\cite{saito2016gate}. These experimental measurements verified that the SC of the ion-gated bulk black phosphorus can reach 510~$\mu$V/K at 210~K which may result in  
a great increase of $ZT$ compared to the bulk single crystal at room temperature.  
The puckered structure of monolayer phosphorene has led to anisotropy in its electrical and thermal properties~\cite{carvalho2016phosphorene}. Theoretical  
studies showed that this distinct feature gives rise to orthogonal electrical and thermal conductances~\cite{n21}. 
This results in a higher $\sigma/\kappa$ ratio and thus the rather large figure of merit $ZT\sim$1 (at room temperature) along the armchair direction of monolayer phosphorene~\cite{n21}. 
Moreover, it is traditionally believed that the utilization of the nanoribbons of such 2D materials is a way to enhance the TE efficiency.  Theoretical calculations predicted that using phosphorene nanoribbons (PNRs) as thermoelectric material can lead to improved TE efficiencies~\cite{n23}. In addition, the effect of the edge states on the thermopower in 
zigzag phosphorene nanoribbons (ZPNRs) was studied~\cite{n25}. It was shown that by applying a transverse electric field, one can completely push the edge modes into the bulk bands and maximize the bulk energy gap which results
in enhanced thermoelectric power in PNRs~\cite{n25}.
Also, it was found that the passivation of edge phosphorous atoms with hydrogen in both types of nanoribbon edges~\cite{n23}, and even oxidation in phosphorene oxide~\cite{n46} enhances the thermopower.\\
These studies confirm that nanoribbon-based
structures of monolayer phosphorene are likely candidates to 
improve the TE properties and the advent of efficient TE nanodevices.
Recent studies reported the production of high-quality PNRs with relatively large and uniform lengths  that may renew the interest in the study of 
thermoelectric properties in nanoribbons and nanostructured systems made of phosphorene~\cite{n22}.
Nanorings made of two dimensional materials are possible nanostructures that may be used to improve TE efficiency. For example, TE properties of rectangular graphene nanorings have been investigated for different conditions and are proposed to build tunable TE generators~\cite{n27,n28}. In such systems, the improvement in the TE performance is attributed to 
the emergence of Fano line shapes or Breit-Wigner line shapes in the 
transmission coefficient that their behavior depends on the geometrical characteristics of the ring and the applied side-gate voltage~\cite{n28}.
 However, despite the numerous works~\cite{saito2016gate,carvalho2016phosphorene,n21,n23,n25,n46} on the TE properties of phosphorene and phosphorene nanoribbons, the TE performance of rectangular phosphorene nanorings has remained elusive so far.\\
In the present work, inspired by the mentioned methods to improve the TE performance in graphene nanorings, we investigate the thermoelectric characteristics of rectangular phosphorene nanorings
for both symmetrically and asymmetrically attachment to the leads. We investigate the effect of an applied external magnetic field
on the TE properties of these rectangular rings and show how the thermoelectric coefficients are controllable in our designed systems by 
 the magnetic field.  Also, we examine the role of ring size and the symmetry of the connection configuration of leads on the TE response of these systems.  Our results will reveal the desirability of these phosphorene-based nanostructures in designing efficient TE devices.
\section{\label{model} MODEL AND FORMALISM }
\begin{figure}[t]
\centering
\includegraphics[width=.45\textwidth]{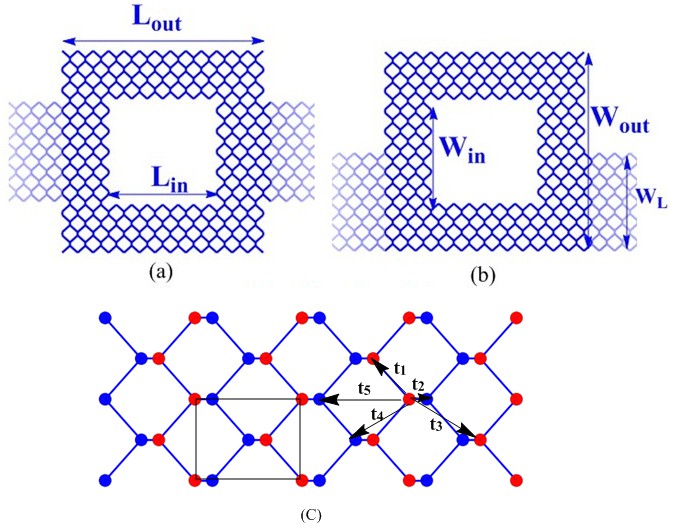}
\caption{ (a) and (b) show the schematic of symmetric  and   asymmetric coupling of a rectangular phosphorene ring to two leads. The geometrical parameters of the system are shown (c) Lattice structure of monolayer phosphorene (top view). There are four P atoms in
the rectangular primitive unit cell of monolayer phosphorene. Red and blue balls denote the puckered structure of the system. $t_1$-t$_5$ denote the used hopping parameters in our TB model.}
\label{aaa}
\end{figure} 
The system under study is a rectangular phosphorene nanoring with a symmetric or asymmetric coupling to two leads as shown in Figs.~\ref{aaa}(a) and (b).  In each configuration, two semi-infinite zigzag- or armchair-edged nanoribbon of phosphorene connect to a nanoring.   
The geometry of a rectangular quantum ring is determined
by the inner and outer sides L$_{in}$, W$_{in}$, L$_{out}$, and W$_{out}$. 
 Ma et al.~\cite{n25} showed that by adjusting the bias voltage of a zigzag phosphorene nanoribbon, the midgap bands completely split and the corresponding localized states give rise to an enhancement in the thermopower. We connect two ZPNRs to the phosphorene nanoring to show that a wide band gap can be induced. This is accompanied by a drastic increase in the magnitude of thermopower compared with the case without the nanoring and even compared with graphene nanorings.
In monolayer phosphorene (MLP), each phosphorus atom (P) is
covalently bonded to three neighbor phosphorus atoms
and form a puckered structure as depicted in
Fig.~\ref{aaa}~(c). The lattice constants of MLP are $a = 4.38$\AA, and $b = 3.31$\AA. There are four P atoms in the rectangular primitive unit cell of MLP.
 The generally accepted low-energy tight-binding (TB) Hamiltonian for MLP is given by~\cite{n32} 
 \begin{equation}
H_c =\sum_{i \neq j}t_{i  j} c_{i}^\dagger c_{j},
\label{eqn:n1}
\end{equation}
 which can also be used to construct the Hamiltonian of PNRs with bare edges.
By applying a perpendicular magnetic field $B$ to the plane of a MLP nanoring, the hopping parameters $t_{i j}$ in Eq.~(\ref{eqn:n1}) are substituted by 
\begin{equation}
t_{i j} \rightarrow t_{i j}\exp\left( i\frac{2\pi e}{h}\int _{\bf{r_i}}^{\bf{r_j}}  \bf{A}\cdot \bf{dl} \right) ,
\label{eqn:n2}
\end{equation}
to include Peierl’s phase factor, where $\bf{A}$ is the vector potential.
To calculate the quantum conductance of a MLP nanoring, we utilize the Landauer-Buttiker formalism as implemented in the KWANT~\cite{n36} package. In this approach, the ballistic transport of charge carriers through a nanoring  is given by
\begin{equation}
G= \frac{2e^2}{h}\int \left(-\frac{\partial f_0}{\partial E}\right) T_{LR}(E) dE.  
\label{eqn:n3}
\end{equation}
In this equation, $f_0$ is the Fermi distribution function $1/ [e^{(E-E_f)/k_BT}+1]$  at temperature $T$, and $T_{LR}(E)$ denotes the the transmission coefficient of our lead-nanoring-lead system~\cite{n33,n34}.
The thermopower $S$ and the electronic contribution to thermal conductance $\kappa_e$  are expressed in terms of the moments of the transmission coefficient~\cite{n35} 
\begin{equation}
K_n= \frac{2}{h}\int _{-\infty}^{\infty} \ \left(-\frac{\partial f_0}{\partial E}\right) (E-\mu)^n T_{LR}(E) dE,  
\label{eqn:n4}
\end{equation}
via the following equations
\begin{equation}
S=- \frac{1}{eT}\frac{K_1}{K_0},  
\label{eqn:n5}
\end{equation}

\begin{equation}
\kappa_e= \frac{1}{T}\left[K_2-\frac{K_1^2}{K_0}\right]. 
\label{eqn:n6}
\end{equation}
To construct the TB models in our systems, we employed the Pybinding code package~\cite{n37}. 
\section{\label{results} RESULTS }
\begin{figure}
\centering
\includegraphics[width=.46\textwidth]{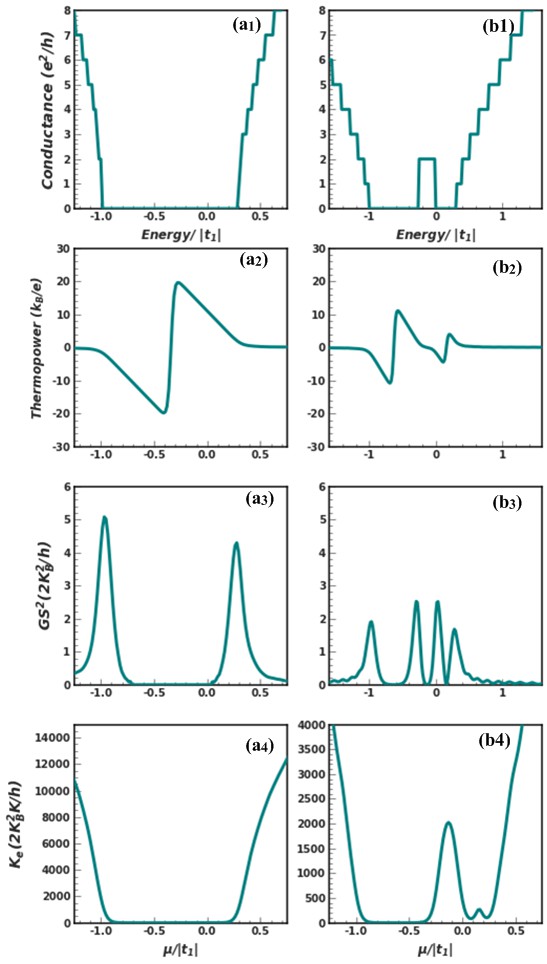}
\caption{ (a1)-(a4) electronic conductance, thermopower, electronic thermal
conductance, and  power factor a function of chemical potential for a APNR with $W=8$~nm. The temperature is set  to $k_BT = 0.03|t_1|$ in our calculations. (b1)-(b4) the same quantities for a ZPNR with $W=8$~nm and at $k_BT = 0.03|t_1|$. }
\label{fig2}
\end{figure}

We design our phosphorene-based nanostructures to enhance the TE performances compared to the TE responses of pristine PNRs.  So, before  proceeding with our main systems,  we first examine the electronic conductance $G$, the Seebeck coefficient $S$,  the electronic thermal conductance $\kappa_e$, and the thermoelectric power factor PF=$GS^2$ for pristine PNRs.  Figures~\ref{fig2}(a1)-(a2) and \ref{fig2}(b1)-(b2) show the calculated conductance and thermopower for a typical armchair phosphorene nanoribbon (APNR) and ZPNR, respectively. We set the width of ribbons to $W=8$~nm for both APNRs and ZPNRs.
The intriguing electronic properties of APNRs and ZPNRs have  a direct consequence on their transport characteristics.
As shown in Figs.~\ref{fig2}(a1) and (b1), the
conductance of both APNRs and ZPNRs exhibit quantized plateaus $G=ne^2/h$, where $e^2/h$ is the conductance quantum and $n$ denotes the number of available transport modes at energy $E$. As seen,  the ZPNR shows a conductance plateau of value $G=2e^2/h$
near the zero energy region because of the contribution of edge-propagating states along the zigzag boundaries. On the other hand,  due to the absence of propagating edge modes in the APNR, there exists no quantized plateau in the midgap region.
\begin{figure}
\centering
\includegraphics[width=.46\textwidth]{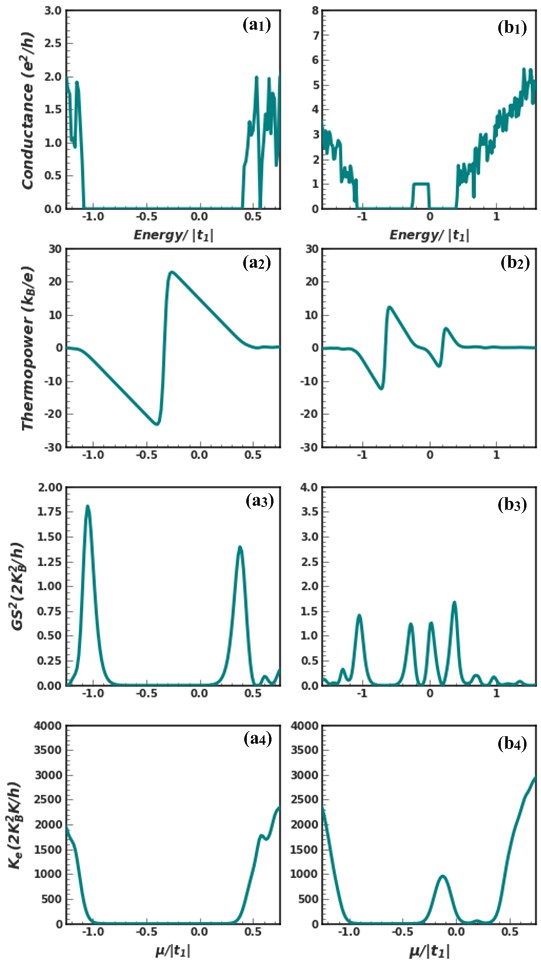}
\caption{ (a1)-(a4) electronic conductance, thermopower, electronic thermal
conductance, and  power factor a function of chemical potential for a symmetric PRZL. The parameters are set to $k_BT = 0.03 |t_1|$, $B=0$, $W_{in}=L_{in}=W_{out}/2=L_{out}/2=W_L=8$~nm.   (b1)-(b4) the same quantities for an asymmetric PRZL with the same set of parameters.} 
\label{fig3}
\end{figure}
\begin{figure}
\centering
\includegraphics[width=.33\textwidth]{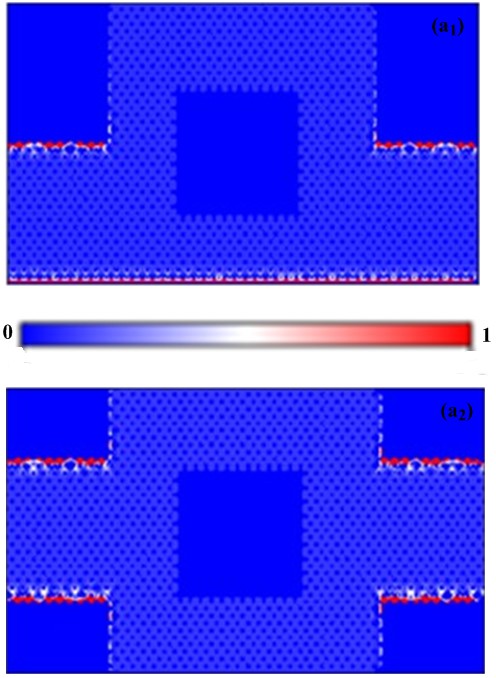}
\caption{Spatial LDOS for (a) a symmetric and  (b) an asymmetric PRZL at $E_f = 0~eV$. The color bar denotes the electronic density distributions across the device.} 
\label{fig4}
\end{figure}
 Furthermore,  the lack of  electron-hole symmetry  in the electronic band structures of PNRs leads to an asymmetry of their quantum conductance  with respect to $E=0$.
 Figures~\ref{fig2}(a2) and (b2) depict the numerically calculated thermopower $S$.  A comparison between the thermopower of the APNR and ZPNR shows that the presence of 
 mid-gap edge modes drastically affects the SC, and hence it is sensitive to the topology of the boundaries in a PNR.  We have shown in Figs.~\ref{fig2}(a3) and (b3) the peaks of the power factor for the condition in which chemical potential sets close to the transmission gap. Also,  Figs.~\ref{fig2}(a4) and (b4) display the electronic 
 thermal conductance $\kappa_e$ as a function of the chemical potential $\mu$.  As seen,  due to the lack of  electron-hole symmetry,
$\kappa_e$ is also asymmetric around
$\mu=0$.
To enhance the SC in phosphorene, one way is to induce a large band gap in phosphorene nanostructures with zigzag edges that have metallic characteristic.  This can be achieved by connecting two narrow leads of PNR to a rectangular phosphorene nanoring as schematically depicted in Fig.~\ref{aaa}. We couple two PNRs with zigzag edges to a rectangular ring to present the mentioned quantities for both symmetric (Figs.~\ref{fig3}(a1)-(a4)) and asymmetric (Figs.~\ref{fig3}(b1)-(b4)) configurations.  
It is found that for a  symmetric connection configuration, the conductance and thermopower are considerably affected compared to the case of pristine PNRs. A comparison between the SCs in Figs.~\ref{fig3}(a2) and ~\ref{fig2}(b2) shows that in this case, the absolute value of the maxima is notably greater than that of the ZPNR counterpart. By comparing Figs.~\ref{fig2}(b1) and ~\ref{fig3}(a1), one can see that for the symmetric case, the first conductance plateau in ZPNR disappears completely,  which is understood as an induced large band gap in the electronic conductance due to the suppression of the
contribution of edge states. We have shown in Fig.~\ref{fig4}(a1) the electronic local density of states (LDOS) for a symmetric PRZL  at $E=0$~eV to clearly show the suppression of the edge modes contribution to the electronic conductance near connection regions. As a result, the peak value of the thermopower  for this PRZL around zero energy is enhanced to 
$23~k_B/e$ ($1978~\mu$V/K).
On the other hand, for an asymmetric PRZL, the situation is different. Figure.~\ref{fig4}(a2) displays the LDOS of electrons for this configuration at  $E=0$~eV. As seen, in this ring-like geometry a new path between the electrodes is observed and the original first conductance plateau still exists (see Figs.~\ref{fig3}(b1)~\ref{fig2}(b1)), though its height decreases due to blocking the other path of edge modes propagation.  To account for these observations, note that for the asymmetric PRZL (symmetric PRZL), the high-energy modes in the terminal are mixed with the high-energy (low-energy) modes in the middle zigzag PNRs (middle
armchair PNRs).  Therefore, the transporting modes are
more significantly affected by scattering for the symmetric than for asymmetric nanorings~\cite{n41}.
Although the maximum power factor in Fig.~\ref{fig2}(b3) is about the same as the one in Fig.~\ref{fig3}(a3), the crucial point here is that by making such nanorings out, we observe a much less electronic thermal conductance $\kappa_e$ (see Fig.~\ref{fig3}(a4)) compared to ZPNRs.  Therefore, our designed phosphorene-based nanostrutre can remarkably contribute to the improvement of the figure of merit $ZT$.

\begin{figure}
\centering
\includegraphics[width=.46\textwidth]{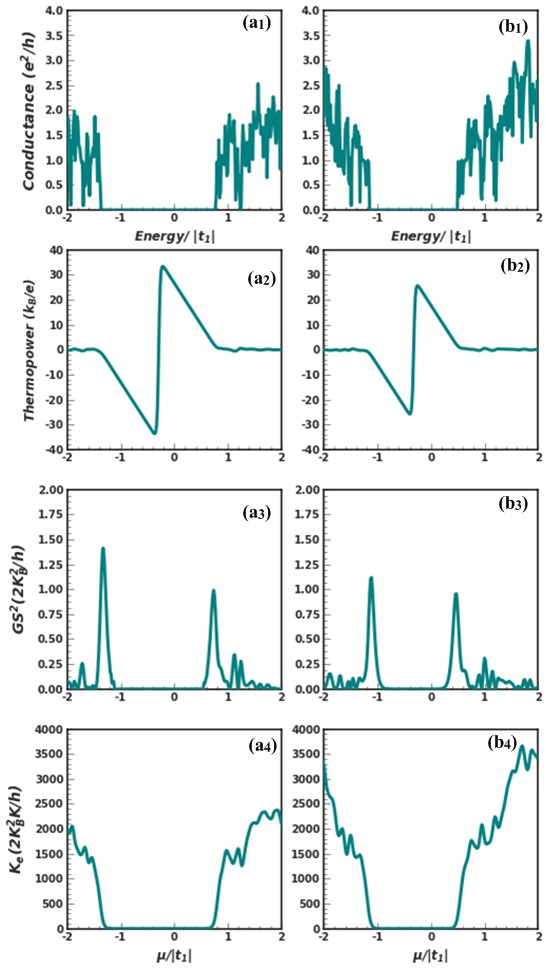}
\caption{(a1)-(a4) electronic conductance, thermopower, electronic thermal
conductance, and  power factor a function of chemical potential for a symmetric PRZL with set of parameters $k_BT = 0.03|t_1|$, $B=0$, $W_{in}=2$~nm, $L_{in}=74$~nm, $W_{out}=8$~nm, $L_{out}=80$~nm, $W_L=4$~nm.   (b1)-(b4) the same quantities for another symmetric PRZL with the set of parameters $k_BT=0.03|t_1|$, $B=0$, $W_{in}=L_{in}=W_{out}/2=L_{out}/2=W_L=4$~nm.}
\label{fig5}
\end{figure}

Next, we examine the impact of the nanoring size on the
electronic conductance and thermopower of PRZLs.
We focus on the symmetric configuration. Fig.~\ref{fig5} depicts 
the calculated conductance, SC, power factor, and electronic thermal
conductance of two PRZLs with different sets of structure parameters. 
Figs.~\ref{fig5}(a1)-(a4) and (b1)-(b4) show the results  for the first set of parameters
$W_{in}=2$~nm, $L_{in}=74$~nm, $W_{out}=8$~nm, $L_{out}=80$~nm, and $W_L=4$~nm, and the 
second set of structure parameters $W_{in}=L_{in}=W_{out}/2=L_{out}/2=W_L=4$~nm, respectively.
The results clearly show that the thermopower profoundly depends 
on the size of considered nanorings.  In general, the smaller ring we consider, the better TE response 
we get. So, in the second case, the maxima $S$ enhances and leads to a  peak value of $33~k_B/e$ ($2830~\mu$V/K) a the zero energy.
\begin{figure}
\centering
\includegraphics[width=.33\textwidth]{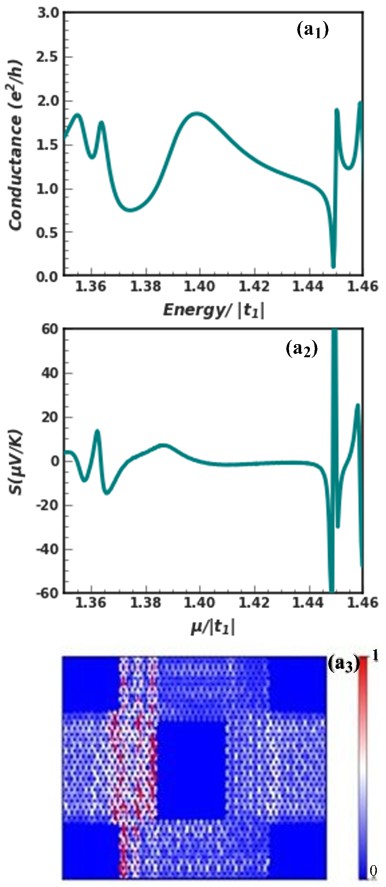}
\caption{ (a1) and (a2) are zoomed-in views of 
Figs.~\ref{fig5}(b1) and (b2) in the energy range of 1.65 to 1.78~eV, respectively. (a3) LDOS of the Fano resonance peak.} 
\label{fig6}
\end{figure}

In addition, the  SC spectrum of the PRZL exhibits some
other small oscillations at higher chemical potentials
yielding changes in the sign. Figures~\ref{fig6}(a1) and (a2) show a zoomed-in view of 
Figs.~\ref{fig5}(b1) and (b2) in the energy range of 1.65 to 1.78~eV. As seen, the 
transmission coefficient shows Fano line shapes that do not exist in ZPNRs.
We have shown in Fig.~\ref{fig6}(a3) the LDOS of the Fano resonance peak. The reason behind the appearance of
Fano resonance peaks is the interaction between the electronic states of the ring and connected leads.
 As seen, the mentioned LDOS shows localized states in some regions of the ring implying that 
 the Fano resonances are closely related to the appearance of these bound states~\cite{n42}.
 Interestingly, when the chemical potential is tuned near 
 a Fano resonance in the transmission, an enhancement in the 
 thermopower of the nanoring is observed. 
 This is justified as follows. 
We suppose that the chemical potential locates near
a Fano-like antiresonance. According to Eq.~(\ref{eqn:n5}), 
the thermopower $S$ is proportional to $K_1$ whose 
value is assessed by the integrand $(-\frac{\partial f_0}{\partial E}) (E-\mu)T_{LR}(E)$ where $(E-\mu)$ 
is an odd function around $\mu$. As a result, a larger 
asymmetry of transmission coefficient $T_{LR}(E)$
leads to an increase in $K_1$, and thus a higher TE response.

Now, let us examine the effect of applying a perpendicular magnetic field on the TE performance of a PRZL. It has been shown~\cite{n44} that by applying a perpendicular magnetic field to PNRs,  the conductance experiences dramatic oscillations at special Fermi energies, leading to a giant magnetoresistance (MR) in zigzag PNRs. Here, we show that the TE performance of a PRZL highly depends on the magnetic flux.
\begin{figure}
\centering
\includegraphics[width=.40\textwidth]{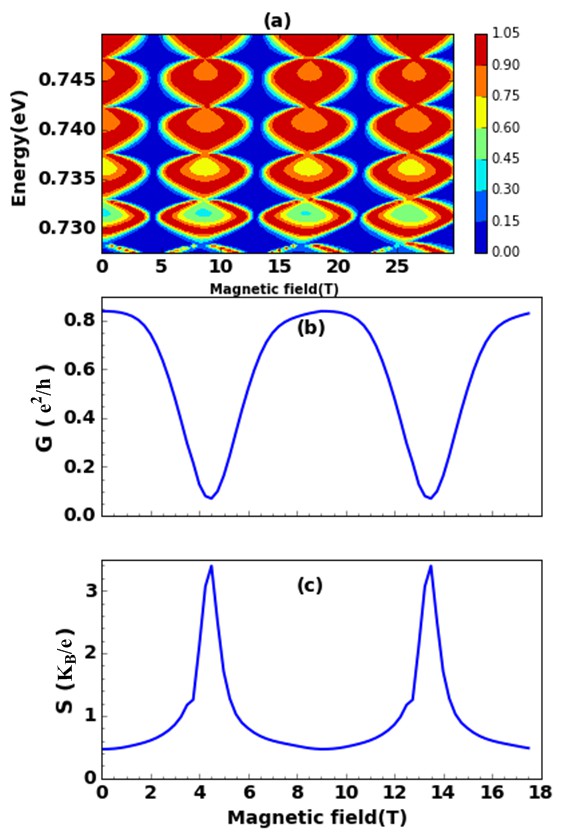}
\caption{ (a) Contour plot of the electronic conductance as a function of the applied magnetic field and the Fermi energy at temperature T = 0 for a symmetric PRZL. The set of parameters are $k_BT = 0.03 |t_1|$, $W_{in}=2$~nm, $L_{in}=74$~nm, $W_{out}=8$~nm, $L_{out}=80$~nm, and $W_L=4$~nm. (b) The electronic conductance of the symmetric PRZL at $E_f=0.732$~eV and $T=0$.} 
\label{fig7}
\end{figure} 
We only focus on the symmetric case because in this configuration the thermopower changes more rapidly compared to the ZPNRs.
Fig.~\ref{fig7}(a) depicts the contour plot of the conductance
as a function of the applied  magnetic field and the Fermi energy  for  a symmetric PRZL with structure parameters $W_{in}=2$~nm, $L_{in}=74$~nm, $W_{out}=8$~nm, $L_{out}=80$~nm, and $W_L=4$~nm.
Note that here we have increased the size of the nanoring to reduce its oscillation period in the presence of the magnetic field. In this situation, the effective area of the nanoring is relatively large, and as a result, the period of AB oscillations is small which makes it experimentally feasible.
We approximate the period of oscillation as $\Delta B=2\pi\phi_0/\bar{S}$, where $\bar{S}=(S_{in}+S_{out})/2$ is the 
average area of the outer and inner nanorings. In our case, the average area is $472.41~nm^2$, yielding the oscillation period of   $\Delta B\thickapprox 8.73~T$.  As seen in Fig.~\ref{fig7}(a), the  
conductance exhibits an oscillative behavior as a function of the applied magnetic field with a period of $\sim8.7$~T which agrees very well with our theoretical prediction.
As seen from Figs.~\ref{fig7}(b) and (c), the electronic conductance $G$ and the thermopower $S$ also have a periodic character by changing the magnetic field. This leads to a drastic increase in the thermopower $S$ near an antiresonance point.
Two salient features that we realize from these figures are the high tunability of the thermopower and the possibility to switch on and off the thermoelectric response of the phosphorene nanoring with the magnetic flux.
 Remarkably, the differential SC can be about eight times larger than the one in the absence of the magnetic field. This confirms the excellent thermoelectric response of this phase-coherent mesoscopic device, making it a promising quantum heat engine in a closed-circuit configuration~\cite{n45}.
\begin{figure}
\centering
\includegraphics[width=.46\textwidth]{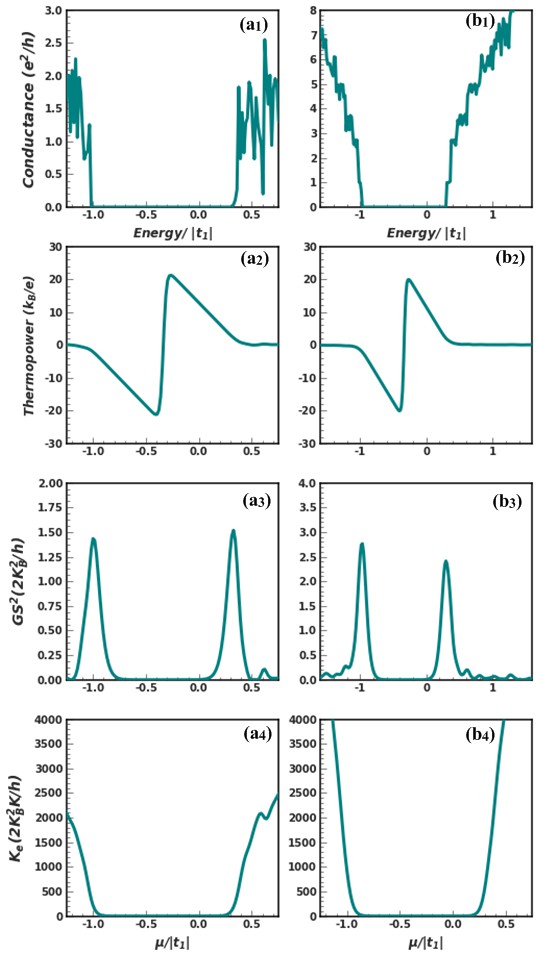}
\caption{ The same quantities as in Fig.~\ref{fig3}  for both (a1)-(a4) symmetric  and (b1)-(b4) asymmetric coupling of two armchair leads two the ring.} 
\label{fig8}
\end{figure}

\begin{figure}[!]
\centering
\includegraphics[width=.35\textwidth]{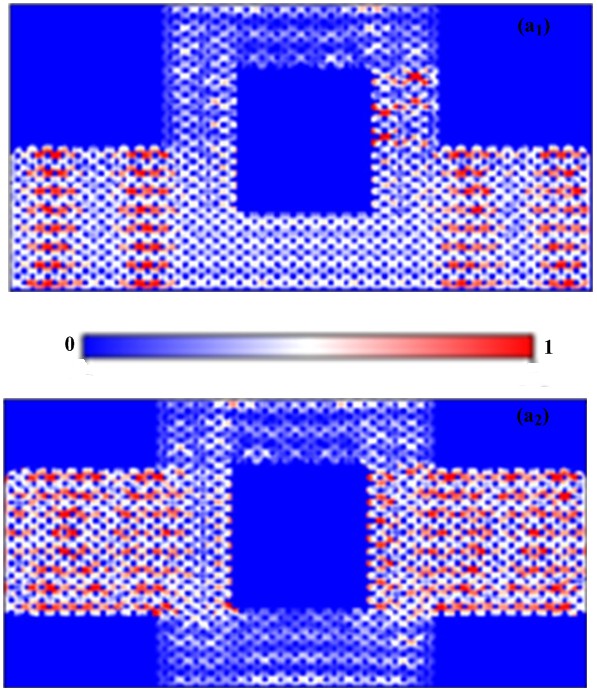}
\caption{ Spatial LDOS for (a) a symmetric and (b) an asymmetric PRAL at $E_f=2$~eV.} 
\label{fig9}
\end{figure}

Finally, we examine the thermoelectric performance of a phosphorene ring with armchair leads (PRAL). Figure~\ref{fig8} depicts the same quantities as in Fig.~\ref{fig3}  for both symmetric ((a1)-(a4)) and asymmetric ((b1)-(b4)) connection configurations. A comparison between the two nanostructures shows that for PRALs, the conductance increases in the asymmetric case. The reason for this observation is that for the asymmetric PRAL,  the high-energy modes in the terminal are mixed with the high-energy modes of the middle armchair PNRs. This phenomenon occurs oppositely for a quantum dot ring geometry.
To further clarify our statement, we have shown in Fig.~\ref{fig9} the spatial LDOS of the transport modes at $E=2$~eV. As seen, in the symmetric case (Fig.~\ref{fig9} (a2)),  the wave packets of electrons are mainly bounded in the attached left and right leads, and one can find fewer bound states in the central device. 
This reduces the propagation probability of electrons from the left lead to the right one or vice versa. On the other hand,  the LDOS of the asymmetric configuration exhibits more bounding  states in
 the central device, and thus increase the probability of electron transmission which emerges as conductance peaks.
The SCs for both the symmetric and asymmetric  PRALs are
about the same as that of APNRs.  Here, the reason is that we have no suppressed edge modes in the ribbons, and thus the band gap remains almost intact.  Intriguingly,  for  PRALs the electronic thermal conductance (see Figs.~\ref{fig8}(a4) and (b4)) can be about ten times 
smaller than that of APNRs (see Fig.~\ref{fig2}(a4)).  As a result, one expects 
to observe a remarkable improvement in the figure of merit ZT for these phosphorene-based nanostructures. The behavior of PRALs by changing the size of the ring and by applying the perpendicular magnetic field is similar to PRZLs and we do not present the results here.
\section{\label{summary} summary and conclusion }
In summary, we investigated the thermoelectric properties of different 
phosphorene nanorings with symmetrical or asymmetrical connection configurations. We utilized the effective low-energy  TB model of monolayer phosphorene to construct the TB Hamiltonian of our designed devices and to characterize their electronic conductance, 
 thermopower, and thermal conductance within the Landauer-Buttiker formalism. We first examined the TE performance of phosphorene nanoribbons to validate our model by comparison with
 other works and then studied the conductance and TE properties of 
our systems. 
Our results showed that the quantum interference
of localized electronic states in the nanorings and the electronic wave packets of leads profoundly affect the TE properties of the system.
We found that for a symmetrical connection of zigzag leads to rectangular rings, the original first conductance plateau completely collapses due to the suppression of the contribution of edge states. 
This induces a wide gap in the system, giving rise to dramatically enhanced peak values in thermopower of such configurations. We also have shown that the TE performance of these systems depends on the size of the ring and one can reach the maximum amount of $33k_B/e$ (2830 $\mu$~V/K) for the thermopower. The appeared Fano antiresonances in the quantum conductance of phosphorene nanorings lead to characteristic features in the Seeback coefficient. We showed the tunability of the thermopower and the possibility to switch on and off the thermoelectric response of phosphorene nanorings by controlling the applied magnetic field. For nanorings with attached armchair phosphorene nanoribbons,  our calculations revealed that though the thermopower almost remains intact, a remarkable decrease in the electronic conductivity occurs.  This can lead to a notable improvement in the figure of merit of these systems. 
Our study confirmed the promising thermoelectric properties of these phase-coherent mesoscopic devices and propose them as potential thermoelectric candidates.
\nocite{*}
\bibliography{phosphorene}

\end{document}